\documentclass{article}
\usepackage[margin=3cm]{geometry}

\usepackage{authblk}

\usepackage{amsmath,amssymb,amsfonts}

\usepackage{makecell}
\setcellgapes{1ex}
\usepackage{booktabs} 
\usepackage{multirow}

\usepackage{graphicx,color}
\usepackage[nocompress]{cite}

\usepackage[shortlabels]{enumitem}

\numberwithin{equation}{section}

\usepackage{hyperref}
\hypersetup{colorlinks=true,allcolors=blue,urlcolor=cyan}

\title{Energy conditions in consistent perfect fluid cosmology}

\author[1]{Davide Batic\footnote{Email: davide.batic@ku.ac.ae}}
\author[2,3]{Christian G. B\"ohmer\footnote{Email: c.boehmer@ucl.ac.uk}}
\author[1]{Denys Dutykh\footnote{Email: denys.dutykh@ku.ac.ae}}

\affil[1]{Department of Mathematics, Khalifa University of Science and \authorcr Technology, PO Box 127788, Abu Dhabi, United Arab Emirates\medskip}
\affil[2]{Department of Mathematics, University College London, \authorcr Gower Street, London WC1E 6BT, UK\medskip}
\affil[3]{Astrophysics Research Centre, School of Mathematics, \authorcr Statistics and Computer Science, University of KwaZulu-Natal, \authorcr Private Bag X54001, Durban 4000, South Africa\medskip}

\date{14 May 2026} 

\pagecolor{white}

\begin{document}

\renewcommand{\arraystretch}{1.2} 
\setlength{\tabcolsep}{1ex} 
\setlength{\extrarowheight}{1ex} 

\maketitle

\begin{abstract}
Motivated by recent work on consistent fluid couplings in $f(R, T)$ gravity, we study cosmology in the nontrivial model $f(R, T) = R + \sigma R T$ using the Brown variational principle for a barotropic perfect fluid. For a flat FLRW universe, we cast the field equations into Einstein-like form and obtain explicit expressions for the effective energy density, pressure and equation of state (EOS) parameter. This allows us to rewrite the null, weak, strong and dominant energy conditions as simple polynomial inequalities. We show that radiation reproduces standard relativistic cosmology, whereas for dust and $\sigma>0$ the effective fluid acquires negative pressure and can drive accelerated expansion. In this dust case, there exists a finite window in the Hubble parameter during which the strong energy condition is violated, but the null, weak, and dominant energy conditions remain satisfied. Conversely, whenever the strong energy condition is imposed, the other conditions are automatically fulfilled. The additional viability requirement $1 + \sigma T > 0$ further restricts the allowed Hubble range and yields an upper bound on $\sigma$ that still leaves a non-empty accelerating regime. Our analysis provides a transparent energy-condition study of a consistent $R\, T$ coupling in $f(R, T)$ cosmology, based on qualitative techniques.
\end{abstract}

\clearpage

\section{Introduction}

Observations of distant type Ia supernovae first revealed that the cosmic expansion is accelerating rather than decelerating, pointing to a dominant component with negative effective pressure in the late universe~\cite{Riess1998AJ, Perlmutter1999AJ}. This picture has been reinforced by high‑precision measurements of the cosmic microwave background (CMB) anisotropies, large‑scale structure and baryon acoustic oscillations, all of which are well described by the spatially flat $\Lambda$CDM model with a cosmological constant and cold dark matter~\cite{Planck2018, Baumann2022}.  In addition to these classic probes, recent large-scale structure surveys such as the Dark Energy Survey (DES) and the Dark Energy Spectroscopic Instrument (DESI) are providing high-precision measurements of weak lensing, galaxy clustering and baryon acoustic oscillations, which further tighten constraints on the late-time expansion history and the dark energy equation of state~\cite{Adame2025JCAP, Abbott2022PRD, Amon2022PRD, Adame2025JCAPa}. Their results significantly reduce the parameter space available for deviations from $\Lambda$CDM and provide stringent targets for any modified-gravity explanation of cosmic acceleration.

Despite this empirical success, the extreme smallness of the observed vacuum energy scale and the absence of a compelling microphysical explanation for $\Lambda$ keep open the possibility that cosmic acceleration may instead reflect a breakdown of general relativity (GR) on cosmological scales~\cite{Weinberg1989RMP, Trodden2011GRG}. A large class of alternative models is provided by modified gravity theories, including $f(R)$ models~\cite{Sotiriou2010RMP}, scalar–tensor theories~\cite{Fujii2009, Capoziello2011PR}, and frameworks with explicit curvature–matter couplings. One of the simplest curvature–matter couplings based on scalar quantities is $f(R, T)$ gravity~\cite{Harko2011PRD}, where the gravitational Lagrangian is an arbitrary function of the Ricci scalar $R$ and of the trace of the matter energy–momentum tensor $T$. This idea has since been embedded in a broader family of models with generalized curvature–matter couplings, such as $f(R, L_m)$ theories, in which the action depends on both curvature invariants and the matter Lagrangian density~\cite{Harko2010EPJC}. These frameworks have been extensively explored in cosmology and astrophysics with the goal of mimicking dark energy~\cite{Sun2016IJMPD}, modifying structure formation, or altering the properties of compact objects through non-minimal interactions between geometry and matter~\cite{Harko2014G, Carvalho2020EPJC, Lobato2022EPJC, Jaybhaye2022PLB, Batic2024PS, Chen2022PDU}. Cosmological applications of $f(R, L_m)$ models with explicit non-minimal couplings of the form $\sigma RL_m$ have shown that even apparently mild curvature–matter interactions can lead to severe viability issues such as a reduced universe lifetime and negative effective densities~\cite{Batic2024PS}. In~\cite{Bhoemer2025PRD} it was shown that the condition $f_{RT} \neq 0$ defines a model with non-trivial new properties. This further motivates a careful analysis of $R\, T$-type couplings in $f(R, T)$ gravity as the simplest models to be explored. 

Within this broader context, the energy conditions provide a particularly simple and powerful diagnostic of physical viability. In GR, they follow from the Raychaudhuri equation combined with the requirement that gravity be attractive, and they constrain the matter content through the null, weak, strong and dominant energy conditions (NEC, WEC, SEC, DEC)~\cite{Hawking1999}. Their generalisation to modified gravity has been worked out for a variety of theories, including $f(R)$ gravity~\cite{Santos2007PRD} and more general curvature‑based modifications~\cite{Capoziello2014PLB}. In $f(R, T)$ gravity,~\cite{Alvarenga2013JMP} used the Raychaudhuri equation to construct the energy conditions for specific choices such as $f(R, T) = R + 2f(T)$, and obtained bounds on model parameters from the requirement that NEC, WEC and DEC be satisfied in cosmological backgrounds, while similar analyses have recently been extended to more general $f(R, L_m, T)$ theories that depend simultaneously on $R$, the matter Lagrangian $L_m$ and the trace $T$, see for example~\cite{Arora2024EPJP}. These studies highlight how energy conditions can systematically constrain the parameter space of curvature–matter coupling models.

However, perfect fluid couplings in $f(R, T)$ gravity are much more involved than in GR. In theories where the action depends explicitly on $T$, the choice of matter Lagrangian cannot be treated with indifference. While the off‑shell Lagrangians for fluids in GR are equivalent, for example, the two most common choices $L_m = -\rho$ and $L_m = p$, these lead to inequivalent theories once the variation of $T$ with respect to the metric and fluid variables is taken into account properly, see also~\cite{Minazzoli2012PRD} for related perfect fluid Lagrangian issues. A recent systematic analysis based on standard relativistic fluid actions has shown that the extra terms entering the modified Einstein equations in $f(R, T)$ gravity depend sensitively on the off‑shell fluid representation, and that the widely used on‑shell prescriptions can generate spurious contributions and inconsistent conservation equations. In particular, for broad classes of functions including separable models such as $f(R, T)=f_1(R)+f_2(T)$ and couplings linear in $T$, the trace dependence can be completely absorbed into a redefinition of the fluid equation of state, so that no genuinely new gravitational degrees of freedom arise for perfect fluids~\cite{Bhoemer2025PRD}. This implies that many existing $f(R, T)$ cosmologies, including several energy condition studies built on these functional forms, either rely on inconsistent fluid dynamics or reduce to effectively standard $f(R)$ theories when treated consistently. These results single out non‑separable curvature–trace couplings as particularly interesting targets for further exploration. Among the simplest such possibilities are models with an explicit bilinear $R\, T$ mixing term, leading to non-minimal couplings that cannot be removed by redefining the matter sector. $R\, T$‑type couplings have started to attract attention in the early‑universe context, where they have been shown to modify the dynamics of slow‑roll inflation and can be constrained by CMB observables~\cite{Jha2025NPB}. To our knowledge, however, there is still no transparent, fully analytic analysis of homogeneous and isotropic cosmology for the simplest $R\, T$ coupling that simultaneously uses a consistent variational description of a barotropic perfect fluid and implements the standard energy conditions in terms of the resulting effective fluid.

In this work, we address this gap by focusing on the minimal non‑separable model $f(R, T)=R+\sigma R T$, where $\sigma$ is a constant coupling parameter. Since $R$ and $T$ differ in units by $\kappa$, the coupling parameter $\sigma$ should have units of $1/\kappa$ to ensure that both terms in $f$ have the same units. This choice represents the simplest extension of the Einstein–Hilbert action with an irreducible curvature–matter trace interaction that depends on a single new parameter. This model lies outside the equivalence classes that can be mapped back to $f(R)$ gravity with an unusual fluid~\cite{Bhoemer2025PRD}. One can also consider this the simplest non-trivial model since $f_{RT}=\sigma\neq0$. Specialising to a spatially flat Friedmann–Lemaitre-Robertson–Walker (FLRW) universe filled with a barotropic perfect fluid, we employ Brown’s Lagrange multiplier formulation of relativistic fluids and the consistent perfect‑fluid treatment of $f(R, T)$ gravity developed in recent work to recast the modified field equations into Einstein‑like form with an effective energy density, pressure and EOS parameter. This effective‑fluid picture allows us to express the NEC, WEC, SEC and DEC as simple inequalities involving the Hubble parameter and the microscopic EOS parameter $w$, making the impact of the chosen coupling on the cosmological dynamics particularly transparent. Our analysis shows that for radiation ($w = 1/3$) the trace $T$ vanishes, the $R\, T$ coupling becomes dynamically irrelevant and standard relativistic cosmology is exactly recovered, whereas for dust ($w = 0$) a positive coupling $\sigma > 0$ endows the effective fluid with a negative pressure and can drive late time accelerated expansion. In the dust case, we identify a finite window in the Hubble parameter where the Universe accelerates, the SEC is violated, but the NEC, WEC and DEC all remain satisfied. Imposing the additional viability requirement that the effective gravitational coupling remain positive, this means $1 + \sigma T > 0$, further restricts the allowed Hubble range yet still leaves room in the parameter space for accelerating expansion to take place. Finally, by matching the effective EOS of the dust‑like effective fluid to observations of the present expansion history, we obtain an order of magnitude estimate for $\sigma$ and show that a coupling of order of the inverse critical density at present can reproduce the observed late time acceleration without invoking any exotic dark energy components. The price to pay to avoid an explicit dark energy component is the non-minimal matter coupling.

The paper is organised as follows. In Sec. II, we review the general formalism of $f(R, T)$ gravity with a consistent perfect fluid sector and introduce the $f(R, T) = R + \sigma RT$ model in the effective‑fluid language without specialising to any particular EOS. Section III summarises the standard energy conditions in terms of the effective energy density and pressure. In Sec. In IV, we apply this framework to a spatially flat FLRW universe filled with a barotropic fluid, derive closed analytic expressions for the effective cosmological dynamics in the dust and radiation cases, and discuss the corresponding constraints on $\sigma$ and the late-time expansion history. Section V contains our conclusions and an outlook on possible extensions, including the impact of $R\, T$ couplings on relativistic stellar structure and compactness bounds.

\section{General formalism and energy conditions}
\label{Sec2}

In 2011,~\cite{Harko2011PRD} proposed a different type of non-minimal matter coupling, in which the gravitational action depends on the Ricci scalar and on the trace of the energy–momentum tensor, $T=g^{\mu\nu}T_{\mu\nu})$. This theory, now known as $f(R,T)$ gravity, is defined by the action
\begin{equation}
  S = \frac{1}{2\kappa}\int f(R,T)\sqrt{-g}~d^4x + \int L_m \sqrt{-g}~d^4x\quad\kappa=8\pi G
\end{equation}
and quickly became highly influential, with well over $2\,000$ citations and numerous cosmological and astrophysical applications. In their seminal paper,~\cite{Harko2011PRD} derived the metric field equations in terms of the energy–momentum tensor $T_{\mu\nu}$ and the tensor $\Theta_{\mu\nu} \equiv g^{\alpha\beta}\delta T_{\alpha\beta}/\delta g^{\mu\nu}$, and expressed the modified Einstein equations in the compact form
\begin{equation}
  f_R R_{\mu\nu} - \frac12 f g_{\mu\nu} + (g_{\mu\nu}\Box - \nabla_\mu\nabla_\nu)f_R = \kappa T_{\mu\nu} - f_T (T_{\mu\nu} + \Theta_{\mu\nu}),
\end{equation}
with $f_R=\partial f/\partial R$ and $f_T=\partial f/\partial T$.  At this abstract level,~\cite{Bhoemer2025PRD} showed that the derivation is correct and provides a natural starting point for exploring trace couplings for arbitrary matter sources. Difficulties arise when $f(R, T)$ gravity is specialised to relativistic perfect fluids. It is well known that the energy–momentum tensor of a perfect fluid can be obtained from several seemingly different matter Lagrangians that are equivalent on shell, i.e. once the fluid equations of motion are imposed. For instance, one can use $L_m=-\rho\sqrt{-g}$ or $L_m=p\sqrt{-g}$, where $\rho$ and $p$ are the energy density and pressure in the fluid rest frame. In standard general relativity or in pure $f(R)$ gravity, these on–shell choices lead to the same field equations, because $L_m$ enters only through $T_{\mu\nu}$. However, in theories where the Lagrangian depends explicitly on $T$, this equivalence breaks down because the variation of the action involves the variation of $T$ itself, and hence it depends on how the fluid is represented off shell in terms of its fundamental variables.~\cite{Bhoemer2025PRD} proved that using $L_m=-\rho\sqrt{-g}$ or $L_m=p\sqrt{-g}$ as fundamental off–shell Lagrangians is not legitimate once the trace of $T_{\mu\nu}$ appears explicitly in the gravitational part of the action. In~\cite{Harko2011PRD}, the perfect–fluid sector was constructed by inserting the on–shell Lagrangians directly into the action and then applying the general field equation. \begin{equation}
f_R R_{\mu\nu} + (g_{\mu\nu}\Box - \nabla_\mu\nabla_\nu)f_R - \frac12 f g_{\mu\nu}
= \kappa T_{\mu\nu} + f_T(\rho+p)U_\mu U_\nu,
\end{equation}
where $U^\mu$ is the fluid four–velocity. The extra term $f_T(\rho+p)U_\mu U_\nu$ has been widely used in the literature and underlies a large number of cosmological and astrophysical applications of $f(R, T)$ gravity. Nonetheless, a careful off–shell variational treatment of the fluid shows that this term is spurious because the factor $\rho+p$ is replaced by a different combination of thermodynamic derivatives, and the resulting equations of motion differ from those adopted by~\cite{Harko2011PRD}. This issue was addressed systematically in~\cite{Bhoemer2025PRD} by deriving the field equations of $f(R, T)$ gravity for relativistic perfect fluids starting from two standard off–shell fluid actions: Brown’s Lagrange multiplier formulation and Schutz’s velocity–potential formulation. In Brown’s picture, the fluid Lagrangian is written in terms of the particle–number density $n$, entropy per particle $s$, a particle–number flux $J^\mu$, and several Lagrange multipliers, with $\rho=\rho(n,s)$ taken as fundamental. In Schutz’s picture, one uses a Lagrangian expressed in terms of the pressure  $p(\mu,s)$, where $\mu$ is the chemical potential. In both approaches~\cite{Bhoemer2025PRD} performed the full variation of the $f(R, T)$ action. A careful analysis of the variation of $T$ with respect to the metric and the fluid variables led to consistent gravitational field equations and fluid equations of motion. The resulting equations can be written in compact forms
\begin{equation}\label{FE}
  f_R R_{\mu\nu} + (g_{\mu\nu}\Box - \nabla_\mu\nabla_\nu)f_R - \frac{1}{2} f g_{\mu\nu} = \kappa T_{\mu\nu} - \frac{1}{2} f_T n\frac{\partial T}{\partial n}(g_{\mu\nu}+U_\mu U_\nu),
\end{equation}
in Brown’s variables, and
\begin{equation}
  f_R R_{\mu\nu} + (g_{\mu\nu}\Box - \nabla_\mu\nabla_\nu)f_R - \frac{1}{2} f g_{\mu\nu} = \kappa T_{\mu\nu} + \frac{1}{2} f_T \mu\frac{\partial T}{\partial\mu}U_\mu U_\nu,    
\end{equation}
in Schutz’s variables. These expressions differ from the results in~\cite{Harko2011PRD} in the replacement of $\rho+p$ by the derivatives $n\partial T/\partial n$ or $\mu\partial T/\partial\mu$, and lead to a consistent set of conservation equations when combined with the fluid equations of motion. An important outcome of this analysis is the emergence of effective thermodynamic quantities induced by the trace coupling and the appearance of second derivatives of the equation of state through $n\partial T/\partial n$ or $\mu\partial T/\partial\mu$. In the Brown formulation, the additional terms in the field equations can be absorbed into an effective pressure $p_{\rm eff}$, while in the Schutz formulation, they manifest as an effective energy density $\rho_{\rm eff}$. This allows the $f(R, T)$ field equations with a perfect fluid to be rewritten in terms of an effective energy–momentum tensor of perfect–fluid form, with new temperature and chemical–potential contributions arising from the non-minimal coupling. Using this effective–fluid picture,~\cite{Bhoemer2025PRD} showed that for a broad class of models, i.e. those with separable dependence $f(R, T) = f^{(1)}(R) + f^{(2)}(T)$ and those linear in $T$, the trace coupling introduces no genuinely new gravitational degrees of freedom for perfect fluids. Instead, the entire effect can be reinterpreted as a reparametrisation of the fluid EOS and, in some cases, as a redefinition of the fluid variables themselves. In particular, $f(R,T)=f^{(1)}(R)+f^{(2)}(T)$ with a perfect fluid is dynamically equivalent to an $f^{(1)}(R)$ theory with an unusual but otherwise standard matter source. These results have far-reaching implications for the $f(R, T)$ literature. Many cosmological models, stellar structure analyses, and phenomenological reconstructions built on the perfect–fluid sector of~\cite{Harko2011PRD} assume precisely the classes of functions $f(R, T) = f^{(1)}(R) + f^{(2)}(T)$ or $f(R, T) = R + 2\lambda T$, combined with an on–shell fluid Lagrangian and the field equation with the extra term $f_T(\rho+p)U_\mu U_\nu$.~\cite{Bhoemer2025PRD} proved that such models either rely on incorrect field equations or, when reinterpreted with the correct equations, reduce to $f(R)$ gravity with a suitably modified EOS. Consequently, a substantial part of the existing $f(R, T)$ phenomenology must be revisited and, in many cases, reinterpreted. In the case of $R+2\lambda T$, this is simply general relativity with a redefined matter tensor. Throughout this section, we take the perfect fluid energy–momentum tensor to be
\begin{equation}\label{31}
    T_{\mu\nu} = (\rho + p)U_\mu U_\nu + p g_{\mu\nu}, \qquad U_\mu U^\mu = -1 ,    
\end{equation}
with $\rho$ and $p$ the energy density and pressure measured in the fluid rest frame. In order to have a concrete matter model in the cosmological applications treated in Section~\ref{FRT}, we now fix the perfect fluid to be barotropic in the sense of Brown’s variational formalism. In this approach, the fluid energy density $\rho$ is taken as a function of the particle number density $n$ and entropy $s$, while the pressure is a derived quantity. For isentropic flows, $s=\mathrm{const}$, Brown’s formalism gives
\begin{equation}\label{pdef}
    p = n\frac{\partial\rho}{\partial n} - \rho,
\end{equation}
see e.g.~\cite{Brown1993CQG}. To close the system of equations and to specify a concrete matter model, it is necessary at this stage to choose an EOS. We take the standard polytropic form $\rho(n)=n^{1+w}$, which, via~\eqref{pdef}, leads to the linear equation of state $p=w\rho$ relating the pressure to the energy density. For such a fluid, the trace of the energy–momentum tensor $T = -\rho + 3p$ can be written as
\begin{equation}
    T = (3w-1)\rho,
    \qquad
    n\frac{\partial T}{\partial n} = (1+w)(3w-1)\rho.
    \label{T_and_n_dTdn}
\end{equation}
An important special case is dust, $w=0$, for which $\rho=n$, $T=-\rho$, and $n\,\partial T/\partial n = -\rho$, corresponding (up to an overall mass scale) to the usual choice $\rho=n$ for cold matter. Finally, for the model $f(R,T)=R+\sigma RT$, the field equations \eqref{FE} become
\begin{equation}
    \label{FEspecial}
    G_{\mu\nu}=R_{\mu\nu}-\frac{1}{2}Rg_{\mu\nu}=\kappa T^{(\rm eff)}_{\mu\nu}
\end{equation}
with
\begin{equation}\label{Tbrown}
T^{(\rm eff)}_{\mu\nu}= \frac{1}{1+\sigma T}\left\{T_{\mu\nu}-\frac{\sigma}{\kappa}\left[(g_{\mu\nu}\Box - \nabla_\mu\nabla_\nu)T
+\frac{1}{2}R n\frac{\partial T}{\partial n}(g_{\mu\nu}+U_\mu U_\nu)\right]\right\}.
\end{equation}
In \(f(R, T)\) gravity, the decomposition of the modified field equations into an effective fluid is not unique because different fluid formulations yield different, non-equivalent field equations. In Brown's Lagrange multiplier formulation, we adopt the effective energy–momentum tensor defined in~\eqref{Tbrown}, which absorbs all $\sigma$-dependent contributions into the right-hand side of~\eqref{FEspecial}. This choice is convenient for interpreting the curvature–trace corrections as modifications of the matter sector. For comparison, in Schutz's velocity–potential formulation, one finds instead
\begin{equation}
    \label{Teff-Schutz}
    T^{(\rm eff)}_{\mu\nu}
    =\frac{1}{1+\sigma T}
    \left[
    T_{\mu\nu}
    -\frac{\sigma}{\kappa}\left(
    (g_{\mu\nu}\Box-\nabla_\mu\nabla_\nu)T
    +\frac{1}{2} R\,\mu\frac{\partial T}{\partial \mu}\,U_\mu U_\nu
    \right)
    \right],
\end{equation}
where $\mu$ is the chemical potential and $U^\mu$ is the fluid four–velocity. In this case, the additional terms induced by the $RT$ coupling are encoded in the derivative $\mu\,\partial T/\partial\mu$ rather than $n\,\partial T/\partial n$. Independent of the chosen fluid representation, the Einstein–like form of the field equations in~\eqref{FEspecial} can be equivalently written as
\begin{equation}
    \label{Rmn-Teff}
    R_{\mu\nu} = \kappa T^{(\rm eff)}_{\mu\nu}
    -\frac{\kappa}{2} T^{(\rm eff)} g_{\mu\nu},
\end{equation}
with $T^{(\rm eff)} = g^{\alpha\beta}T^{eff)}_{\alpha\beta}$. This form provides the natural starting point for expressing the Raychaudhuri-based energy conditions in terms of the effective fluid in the next section.

Finally, for the model $f(R, T) = R + \sigma R T$ we have $f_R = 1+\sigma T$, so that the Einstein–like form of the field equations involves an effective gravitational coupling $\kappa_{\rm eff} = \kappa/(1+\sigma T)$. In order to avoid a change of sign of $\kappa_{eff}$ and the associated anti‑gravity behaviour, as well as the singular limit $1+\sigma T=0$, we shall restrict attention to backgrounds satisfying $1+\sigma T>0$. This condition is the direct analogue of the standard requirement $f_R>0$ in $f(R)$ gravity and ensures that the effective energy density and pressure appearing in the energy conditions have the usual interpretation. We have $T=-\rho+3p$ and $\rho \gg p$ for standard matter which means $T \approx -\rho$ in this case, so that $\kappa_{\rm eff} \approx \kappa/(1-\sigma\rho)$. This suggests that $\sigma < 0$ would be the more natural choice. This argument can easily be generalised to a more general equation of state.

In order to formulate the energy conditions in $f(R, T)$ gravity, we first recall the Raychaudhuri equation for congruences of timelike and null curves. For a timelike congruence with tangent vector $V^\mu$, the evolution of the expansion scalar $\theta$ along the affine parameter $\tau$ is governed by equation~(8) in~\cite{Alvarenga2013JMP}, while for a null congruence with tangent vector $k^\mu$ and parameter $\lambda$, the corresponding relation is given by their equation (9). In both cases, $\theta$ measures the volume expansion of the congruence, while $\sigma_{\mu\nu}$ is the shear tensor encoding distortions and $\omega_{\mu\nu}$ is the vorticity tensor describing rotations. Under the assumption of negligible shear and vanishing vorticity, the quadratic contributions in $\theta$, $\sigma_{\mu\nu}$ and $\omega_{\mu\nu}$ can be dropped, and integration of the Raychaudhuri equation then yields the approximate relation (10) in~\cite{Alvarenga2013JMP}, which expresses $\theta$ directly in terms of $R_{\mu\nu}V^\mu V^\nu$ and $R_{\mu\nu}k^\mu k^\nu$, respectively. Hence, the strong energy condition (SEC) and the null energy condition (NEC) become
\begin{equation}
  R_{\mu\nu}V^\mu V^\nu\geq 0, \qquad R_{\mu\nu}k^\mu k^\nu\geq 0.
\end{equation}
It is straightforward to verify that they are equivalent to 
\begin{equation}
  \left(T^{(\rm eff)}_{\mu\nu}-\frac{1}{2}T^{(\rm eff)}g_{\mu\nu}\right)V^\mu V^\nu\geq 0,\qquad T^{(\rm eff)}_{\mu\nu}k^\mu k^\nu\geq 0.
\end{equation}
On the other hand, we can also define 
\begin{equation}
    T^{(\rm eff)}_{\mu\nu } = (\rho_{\rm eff}+p_{\rm eff})U_\mu U_\nu+p_{\rm eff}g_{\mu\nu}
\end{equation}
with
\begin{equation}
  \rho_{\rm eff}=T^{(\rm eff)}_{\mu\nu} U^\mu U^\nu, \qquad p_{\rm eff}=\frac{1}{3}h^{\mu\nu}T^{(\rm eff)}_{\mu\nu}
\end{equation}
with a spatial projector
\begin{equation}
  h^{\mu\nu} = g^{\mu\nu} + U^\mu U^\nu.
\end{equation}
With this specification, we can carry over the energy conditions derived by~\cite{Alvarenga2013JMP}. More precisely, we have
\begin{eqnarray}
\mbox{SEC}&:&\quad \rho_{\rm eff}+p_{\rm eff}\geq 0,\qquad \rho_{\rm eff}+3p_{\rm eff}\geq 0,\label{SEC}\\
\mbox{NEC}&:&\quad \rho_{\rm eff}+p_{\rm eff}\geq 0,\label{NEC}\\
\mbox{WEC}&:&\quad \rho_{\rm eff}\geq 0,\qquad \rho_{\rm eff}+p_{\rm eff}\geq 0,\label{WEC}\\
\mbox{DEC}&:&\quad \rho_{\rm eff}\geq 0,\qquad \rho_{\rm eff}+p_{\rm eff}\geq 0,\qquad \rho_{\rm eff}-p_{\rm eff}\geq 0.\label{DEC}\\
\end{eqnarray}
It is worth emphasising that the NEC is not equivalent to $\rho_{\rm eff}\geq 0$. By definition, the NEC follows from the requirement $T^{(\rm eff)}_{\mu\nu}k^\mu k^\nu \geq 0$ for any null vector $k^\mu$. For a perfect fluid energy-momentum tensor, this contraction yields $T^{(\rm eff)}_{\mu\nu}k^\mu k^\nu = (\rho_{\rm eff}+p_{\rm eff})(U_\mu k^\mu)^2$, since $k^\mu k_\mu = 0$. Because $(U_\mu k^\mu)^2>0$, the NEC reduces to the condition $\rho_{\rm eff}+p_{\rm eff}\geq 0$. The inequality $\rho_{\rm eff}\geq 0$ instead belongs to the WEC and cannot, by itself, guarantee the focusing of null geodesics.

\section{Flat Friedman--Robertson--Walker universe with \texorpdfstring{$f(R,T)=R+\sigma RT$}{f(R,T) = R + σ RT}}\label{FRT}

In this section, we determine the conditions under which the energy conditions are satisfied in the present model. To this end, we first establish the explicit expressions for the effective energy density and effective pressure arising from the modified field equations. Let us first recall that the Friedmann-Lemaitre-Robertson-Walker (FLRW) universe is described by the metric
\begin{equation}
  ds^2=-dt^2+a^2(t)(dx^2+dy^2+dz^2),
\end{equation}
where $a(t)$ is the scale factor. We further assume a comoving fluid with 4-velocity $U^\mu = (1,0,0,0)$ while $T$ is described by the homogeneous scalar field $T = T(t)$. A straightforward computation gives
\begin{equation}
  (g_{\mu\nu}\Box-\nabla_\mu\nabla_\nu)T=\left\{
    \begin{array}{cc}
    3H\dot{T} & \mbox{if}~\mu=\nu=0\\
    -g_{ij}(\ddot{T}+2H\dot{T}) &\mbox{if}~\mu=i,~\nu=j
    \end{array}\right.,\qquad H=\frac{\dot{a}}{a},
\end{equation}
and moreover, in our comoving choice $g_{00}+U_0 U_0=0$ and $g_{ij}+U_i U_j=g_{ij}$. Finally, in the FLRW metric, the Ricci scalar is $R=6(\dot{H}^2+2H^2)$ for the spatially flat case $k=0$. The effective energy density is
\begin{equation}
    \label{rhoeff}
    \rho_{\rm eff}=\frac{1}{1+\sigma T}
    \left(\rho-\frac{3\sigma}{\kappa}H\dot{T}\right)
\end{equation}
while the effective pressure reads
\begin{equation}
    \label{peff}
    p_{\rm eff} = \frac{1}{1+\sigma T}\left[p+\frac{\sigma}{\kappa}(\ddot{T}+2H\dot{T})-\frac{3\sigma}{\kappa}(\dot{H}+2H^2)n\frac{\partial T}{\partial n}\right],
    \qquad 
    T=-\rho+3p.    
\end{equation}
Furthermore, we require $1+\sigma T>0$. This condition ensures a positive and well-defined effective gravitational coupling $\kappa_{\mathrm{eff}}=\kappa/(1+\sigma T)$. Moreover, the cosmological field equations are
\begin{equation}
    \label{Friedman}
    3H^2=\kappa\rho_{\rm eff},\qquad 
    2\dot{H}+3H^2=-\kappa p_{\rm eff}.
\end{equation}
In order to close the system of equations and to specify a concrete matter model, we now specialise to the barotropic Brown fluid introduced in Sec. II, with a constant EOS parameter $w$, $\rho(n)=n^{1+w}$, and $p=w\rho$  so that $T$ and $n\partial T/\partial n$ are given by \eqref{T_and_n_dTdn}. 

In Brown's variational formulation, the fundamental fluid equation is the conservation of the vector-density particle-number flux $J^\mu$~\cite{Brown1993CQG},
\begin{equation}\label{Jcons}
  \partial_\mu J^\mu = 0, \qquad J^\mu = \sqrt{-g} nU^\mu.
\end{equation}
For the spatially flat FLRW metric with comoving four-velocity $U^\mu = (1,0,0,0)$ this immediately reduces to
\begin{equation}\label{n_scaling}
    \partial_t\!\left(a^3 n\right) = 0
    \quad \Rightarrow \quad n \propto a^{-3}.
\end{equation}
Using the barotropic equation of state $\rho(n) = n^{1+w}$ introduced in Sec.~\ref{Sec2}, we obtain the usual scaling law $\rho(a) \propto a^{-3(1+w)}$. Differentiating with respect to time and using $p = w\rho$ yields the standard continuity equation
\begin{equation}\label{rho_continuity}
    \dot\rho + 3H(1+w)\rho = 0.
\end{equation}
Thus, in the present $RT$ model, the familiar GR continuity relation is not assumed by hand but follows from particle-number conservation together with the chosen Brown fluid equation of state. Equivalently, as shown in~\cite{Bhoemer2025PRD}, the timelike projection of the modified conservation equations is unchanged and can be written as $U_\nu \nabla_\mu T^{\mu\nu} = 0$, so that~\eqref{rho_continuity} also represents the parallel projection of $\nabla_\mu T^{\mu\nu}$. The non-minimal $RT$ coupling contributes only to the orthogonal projection $h^\sigma{}_\nu \nabla_\mu T^{\mu\nu}$, which describes an energy exchange in the perpendicular direction to the four-velocity and vanishes identically for the homogeneous and isotropic FLRW background. Taking into account that
\begin{equation}
    \dot{T}=3(1+w)(1-3w)H\rho,\quad
    \ddot T =3(1+w)(1-3w)\left[\dot{H}-3(1+w)H^{2}\right]\rho,
\end{equation}
Substituting these relations into \eqref{rhoeff} and \eqref{peff}, and using the Friedmann equations \eqref{Friedman}, the effective energy density and pressure can be written purely in terms of the quantities $\rho$, $H$, $\dot{H}$ and $w$ as
\begin{eqnarray}
    \rho_{\rm eff} &=& \frac{\kappa+9\sigma(1+w)(3w-1)H^2}{\kappa\left[1+\sigma(3w-1)\rho\right]}\rho,
    \label{rho_eff_barotropic}\\
    p_{\rm eff} &=& \frac{\kappa w+3\sigma(3w+2)(1+w)(3w-1)H^2}{\kappa\left[1+\sigma(3w+2)(1-3w)\rho\right]}\rho,
    \label{p_eff_barotropic}
\end{eqnarray}
where we used the second Friedmann equation in \eqref{Friedman} in order to get eliminate the dependence on $\dot{H}$ in \eqref{p_eff_barotropic}. Moreover, the first Friedmann equation in \eqref{Friedman} provides an algebraic relation between the physical density $\rho$ and the Hubble rate
\begin{equation}
    \label{rho_of_H}
    \rho(H) =\frac{3H^2}{\kappa+3\sigma(3w+2)(3w-1)H^2}.
\end{equation}
Combining this algebraic relation with the continuity equation~\eqref{rho_continuity} shows that, as in standard GR, the two Friedmann equations~\eqref{Friedman} are not independent. The matter equation~\eqref{n_scaling} or~\eqref{rho_continuity} together with the first field equations imply the second field equation. The homogeneous background dynamics therefore reduce to a single first–order equation for $H(t)$ (or equivalently $a(t)$), supplemented by the algebraic relation \eqref{rho_of_H} between $H$ and $\rho$. In the remainder of this work, we will not attempt to solve this equation explicitly. Instead, we treat $H$ (or $H^{2}$) as a free parameter constrained only by~\eqref{rho_of_H} and use the effective fluid representation to rewrite the energy conditions as simple polynomial inequalities in $H^{2}$, $w$ and the coupling $\sigma$. This allows a quantitative study of the general properties of the solutions, independently of closed-form solutions which one may not be able to find.

If we replace \eqref{rho_of_H} into \eqref{p_eff_barotropic} we find the useful representation
\begin{equation}\label{peff2}
  p_{\rm eff} = \frac{3H^2}{\kappa^2} \left[\kappa w+3\sigma(3w+2)(3w-1)(1+w)H^2\right].    
\end{equation}
Taking into account that $\rho_{\rm eff}=3H^2/\kappa$, we can introduce an effective EOS parameter defined by
\begin{equation}\label{weff}
    w_{\rm eff}(H)= \frac{p_{\rm eff}}{\rho_{\rm eff}}=w+\frac{3\sigma}{\kappa}(3w+2)(3w-1)(1+w)H^2.
\end{equation}
Notice that $w_{\rm eff}$ depends on the background only through $H$. Moreover, $w_{\rm eff}$ coincides with the bare EOS parameter $w$ in the limit $\sigma\to 0$, as one would expect for models of this type. Furthermore, for $w=1/3$ one has $w_{\rm eff} = 1/3$ and the model reduces exactly to cosmology in the radiation era; see also~\cite{Bhoemer2025PRD}, where this property was stated. For any cosmological model where $H \to 0$ as $t \to \infty$, one finds $w_{\rm eff} \to w$, suggesting that the effects of the trace coupling dominate the early evolution of the Universe. Similarly, if $H \to \lambda$, $0 < \lambda = \text{const.}$, as $t \to \infty$, then one would find
\begin{equation}
    \label{weff2}
    w_{\rm eff} \to w+\frac{3\sigma}{\kappa}(3w+2)(3w-1)(1+w)\lambda^2.
\end{equation}
This is interesting from a phenomenological point of view, since for $\sigma \lambda^2/\kappa \ll 1$ one would obtain $w_{\rm eff} \approx w$ despite the model exhibiting accelerated expansion. At this point, a comment is in order. The parameter $w_{\rm eff}(H) = p_{\rm eff}/\rho_{\rm eff}$ in \eqref{weff} is a convenient way of rewriting the effective stress tensor and the associated energy conditions in algebraic form. However, unlike in standard GR with a barotropic fluid, $w_{\rm eff}$ is not a microscopic EOS constant because it is a derived quantity that depends explicitly on the background through $H(t)$. Consequently, inequalities such as $w_{\rm eff}\ge -1$ or $w_{\rm eff}\ge -1/3$ should be understood as constraints on the values of $H$ along a solution, rather than as independent input assumptions. A simple illustration is provided by assuming a constant value $H_{\rm ds}$ (de Sitter) for the Hubble function. This means we set $a(t) = a_0 \exp(H_{\rm ds} t)$. In this case, $\dot H = 0$ and the Einstein-like Friedmann equations imply
\begin{equation}
    3\lambda^2=\kappa\rho_{\rm eff}, \qquad 
    3\lambda^2=-\kappa p_{\rm eff}
    \qquad\Longrightarrow\qquad
    p_{\rm eff}=-\rho_{\rm eff}
    \quad (w_{\rm eff}=-1).
\end{equation}
Combining $w_{\rm eff}=-1$ with~\eqref{weff} evaluated using $H=H_{\rm ds}$, we arrive at the condition
\begin{equation}
    w + \frac{3\sigma}{\kappa}(3w+2)(3w-1)(1+w)H_{\rm ds}^2 = -1.
\end{equation}
For $w\neq -1$, the condition $w_{\rm eff}(H_{\rm ds})=-1$ places the system on the degeneracy surface of the algebraic relation $\rho(H)$, so the corresponding constant-$H$ point $H_{\rm ds}$ is not a regular FLRW solution with finite matter density (see Appendix~\ref{app:constH}). We emphasise that, once the additional viability requirement $1+\sigma T>0$ is imposed, the exact de Sitter limit in the dust branch lies outside the allowed $H$-range, so that the physically admissible accelerating solutions are quasi-de Sitter rather than exactly de Sitter.

\subsection{Energy conditions}

The standard energy conditions~\eqref{SEC}--\eqref{DEC}, which we originally wrote in terms of the effective energy density $\rho_{\rm eff}$ and effective pressure $p_{\rm eff}$, may be recast in a compact form using the effective equation of state parameter $w_{\rm eff}$, defined in~\eqref{weff}.
\begin{itemize}
\item 
\textbf{NEC}: If we impose \eqref{NEC} together with the EOS $p_{\rm eff}=w_{\rm eff}\rho_{\rm eff}$, the NEC can be expressed as $\rho_{\rm eff}(1+w_{\rm eff}) \geq 0$. Since $\rho_{\rm eff}=3H^2/\kappa>0$ for $H^2>0$, the NEC reduces to $w_{\rm eff} \geq -1$. By means of~\eqref{weff} it becomes
\begin{equation}
    (1+w)\left[1+\frac{3\sigma}{\kappa}(3w+2)(3w-1)H^2\right] \geq 0.
    \label{NEC_inequality}
\end{equation}

\item 
\textbf{WEC}: We need to require $\rho_{\rm eff} \geq 0$ and $\rho_{\rm eff} + p_{\rm eff} \geq 0$. The first inequality is automatically satisfied, while the second coincides with the NEC condition. Hence, in this model, the WEC is equivalent to the NEC.

\item 
\textbf{SEC}: In terms of $w_{\rm eff}$, the conditions $\rho_{\rm eff} + 3p_{\rm eff} \geq 0$ and $\rho_{\rm eff} + p_{\rm eff} \geq 0$ become $w_{\rm eff}\geq -1/3$ and $w_{\rm eff}\geq -1$. Thus, the SEC holds whenever $w_{\rm eff}\geq -1/3$, with the NEC condition included as a subset. The latter inequality can also be cast into the following form
\begin{equation}
    3w+1+\frac{9\sigma}{\kappa}(3w+2)(3w-1)(1+w)H^2\geq 0.    
\end{equation}
Furthermore, violation of the SEC requires $w_{\rm eff} < -1/3$, which is precisely the regime usually associated with accelerated expansion. This can be easily seen from the field equations~\eqref{Friedman} from which we obtain
\begin{equation}
    \frac{\ddot{a}}{a}=-\frac{\kappa}{6}(\rho_{\rm eff}+3p_{\rm eff}).
\end{equation}
Hence, accelerated expansion ($\ddot{a}>0$) is equivalent to $\rho_{\rm eff}+3p_{\rm eff}<0$, i.e.\ to $w_{\rm eff}< -1/3$ along the background solution. In particular, in our effective fluid description, any phase of cosmic acceleration necessarily corresponds to a violation of the strong energy condition.
\item \textbf{DEC}: We need to impose $\rho_{\rm eff} \geq 0$, and $\rho_{\rm eff} \pm p_{\rm eff} \geq 0$. For a perfect fluid, this is equivalent to $-1\leq w_{\rm eff}\leq 1$. Inserting \eqref{weff} yields the equivalent condition
\begin{equation}\label{DEC_inequality}
  -1 \le w + \frac{3\sigma}{\kappa}(3w+2)(3w-1)(1+w)H^2 \le 1.
\end{equation}
\end{itemize}

Table~\ref{tableEinsnone} makes explicit how the sign of the $RT$ coupling and the matter equation of state control the validity of the energy conditions. For radiation ($w=1/3$), all conditions are always satisfied, independently of $\sigma$, reflecting the fact that $T=0$ and the model reduces exactly to cosmology in the radiation era. The dust case ($w = 0$) displays a much richer behaviour. For $\sigma>0$ the effective EOS parameter is $w_{\rm eff}^{(w=0)}(H) = -6\sigma H^2/\kappa$, so that the effective dust fluid acquires a negative pressure and interpolates between $w_{\rm eff}\simeq 0$ at small $H^2$ and $w_{\rm eff}\to-1$ as $H^2$ approaches its upper bound. In this branch, the NEC and WEC coincide and require $H^2\leq\kappa/(6\sigma)$, while the SEC is more restrictive and holds only for $H^2\leq\kappa/(18\sigma)$. Thus, in the intermediate regime $\kappa/(18\sigma)<H^2\leq\kappa/(6\sigma)$, the NEC, WEC, and DEC remain satisfied, but the SEC is violated. This is precisely the regime in which $w_{\rm eff}<-1/3$ and the effective dust component can drive an accelerated expansion. In contrast, for $\sigma<0$ one finds $w_{\rm eff}^{(w=0)}(H)=6|\sigma|H^2/\kappa\geq 0$, so the effective dust becomes stiffer than ordinary dust and can approach a stiff-fluid behaviour ($w_{\rm eff}\to 1$) as $H^2$ grows. In this case, the SEC, NEC and WEC are always satisfied, while the DEC imposes the upper bound $H^2\le\kappa/(6|\sigma|)$ to keep $w_{\rm eff}\leq 1$. No acceleration occurs in this branch, since $w_{\rm eff}\geq 0$ implies $\ddot{a}/a < 0$.

\begin{table}[!hbt]
\centering
\setlength\tabcolsep{0.1cm}
\def\arraystretch{1.5}
\begin{tabular}{ | c | c | c | c | c | c | c | c }
\hline
$\sigma>0$  & $w$  & SEC & NEC & WEC & DEC\\ \hline
            & $0$  & $0\leq H^2\leq \frac{\kappa}{18\sigma}$ & $0\leq H^2\leq \frac{\kappa}{6\sigma}$ & $0\leq H^2\leq \frac{\kappa}{6\sigma}$ & $0\leq H^2\leq \frac{\kappa}{6\sigma}$\\ \hline
            & $1/3$& $\checkmark$ & $\checkmark$ & $\checkmark$ & $\checkmark$\\ \hline \hline
$\sigma<0$  & $w$  & SEC & NEC & WEC & DEC\\ \hline
            & $0$  & $\checkmark$ & $\checkmark$ & $\checkmark$ & $0\leq H^2\leq \frac{\kappa}{6|\sigma|}$\\ \hline
            & $1/3$& $\checkmark$ & $\checkmark$ & $\checkmark$ & $\checkmark$\\ \hline           
\end{tabular}
\caption{Summary of the standard energy conditions in the model $f(R,T)=R+\sigma RT$ for dust ($w=0$), radiation ($w=1/3$), and for both signs of the coupling parameter $\sigma$. Here $\kappa>0$ is the gravitational coupling, $H$ is the Hubble parameter, and the $\checkmark$ indicates that the corresponding condition is satisfied.}
\label{tableEinsnone}
\end{table}

Overall, the table shows that a positive coupling $\sigma > 0$ allows an effective dust component with negative pressure capable of violating the SEC and generating late-time acceleration while preserving NEC, WEC and DEC over a finite range of $H^2$, whereas a negative coupling $\sigma < 0$ yields a stiffer, non-accelerating effective fluid.

In addition to the energy conditions, our model is subject to the requirement $1 + \sigma T > 0$, which plays the same role as $f_R>0$ in $f(R)$ gravity and guarantees a positive effective gravitational coupling $\kappa_{\rm eff}=\kappa/(1+\sigma T)$. For a barotropic fluid with $p=w\rho$ we have $T=(3w-1)\rho$, and in the range of interest $-1<w\leq 1/3$ one has $3w-1\leq 0$ so that $T\leq 0$. The impact of $1+\sigma T>0$ then depends on the sign of the coupling. For $\sigma < 0$, one finds $\sigma T>0$ and hence $1+\sigma T>1$, so the condition is automatically satisfied and does not introduce any additional restriction beyond those already encoded in the energy conditions. For $\sigma>0$, by contrast, the inequality $1+\sigma(3w-1)\rho>0$ implies the upper bound
\begin{equation}\label{ineq}
  \rho < \frac{1}{\sigma(1-3w)}.
\end{equation}
Using the algebraic relation \eqref{rho_of_H} the condition $1+\sigma T>0$, which for $-1<w<1/3$ is equivalent to \eqref{ineq}, can be rewritten as
\begin{equation}\label{ineq1}
  \frac{3H^2}{\kappa-3\sigma(3w+2)(1-3w)H^2} < \frac{1}{\sigma(1-3w)}.
\end{equation}
If we impose $\kappa-3\sigma(3w+2)(1-3w)H^2>0$, which implies
\begin{equation}\label{prebound}
  H^2 < \frac{\kappa}{3\sigma}f_1(w), \qquad f_1(w) = \frac{1}{(3w+2)(1-3w)},
\end{equation}
then the inequality \eqref{ineq1} translates into an additional upper limit on the Hubble rate,
\begin{equation}\label{bound}
  H^2 < H_{\sigma}^2=\frac{\kappa}{3\sigma}f_2(w), \qquad f_2(w)=\frac{1}{3(1+w)(1-3w)}.
\end{equation}
Since $f_2(w) < f_1(w)$ for all $w\in(-1,1/3)$, the upper bound in \eqref{bound} is stronger than \eqref{prebound}, and therefore \eqref{bound} automatically implies \eqref{prebound}. In the same interval one has $H_{\sigma}^2>0$, so the condition $1+\sigma T > 0$ is simply equivalent to
\begin{equation}
  0 \leq H^2\leq H_{\sigma}^2.
\end{equation}
This should be compared with the NEC/WEC/DEC constraint
\begin{equation}
  H^2\leq H_{\mathrm{NEC}}^2 = \frac{\kappa}{3\sigma(3w+2)(1-3w)} = \frac{\kappa}{3\sigma}f_1(w).
\end{equation}
Since $H_{\sigma}^2=(\kappa/3\sigma)f_2(w)$ and $f_2(w)<f_1(w)$, it follows that
\begin{equation}
  H_{\sigma}^2 < H_{\mathrm{NEC}}^2.
\end{equation}
Hence, for $\sigma>0$, the requirement $1+\sigma T>0$ is more restrictive than the null, weak and dominant energy conditions, and must be imposed independently. By contrast, when the SEC holds one has $H^2\leq H_{\mathrm{SEC}}^2$, and
\begin{equation}
  \frac{H_{\mathrm{SEC}}^2}{H_{\sigma}^2} = \frac{3w+1}{3w+2}<1 \qquad (w>-1/3).
\end{equation}
Thus, in the range $w>-1/3$, the SEC bound is the tighter one and already guarantees $1+\sigma T>0$. For dust ($w=0$), the relevant limits are
\begin{equation}
    H_{\sigma}^2 = \frac{\kappa}{9\sigma}, \qquad H_{\mathrm{NEC}}^2 = \frac{\kappa}{6\sigma}, \qquad H_{\mathrm{SEC}}^2 = \frac{\kappa}{18\sigma}.
\end{equation}
Therefore, imposing $1+\sigma T>0$ reduces the NEC/WEC/DEC-allowed interval from
\begin{equation}
  0 \leq H^2 \leq \frac{\kappa}{6\sigma}
\end{equation}
to
\begin{equation}
  0 \leq H^2 \leq \frac{\kappa}{9\sigma},
\end{equation}
while the SEC still requires
\begin{equation}
H^2\leq \frac{\kappa}{18\sigma}.
\end{equation}
The accelerated expansion regime for dust, defined by $w_{\rm eff}<-1/3$ and hence $H^2>\kappa/(18\sigma)$, is therefore not eliminated. It is simply narrowed to
\begin{equation}
  \frac{\kappa}{18\sigma} < H^2\leq \frac{\kappa}{9\sigma}.
\end{equation}
For radiation ($w = 1/3$), one has $T = 0$, so $1 + \sigma T > 0$ is automatically satisfied for any $\sigma$.

\subsection{An estimate for \texorpdfstring{$\sigma$}{σ}}

In order to obtain an indicative magnitude for the curvature-trace coupling $\sigma$, it is natural to appeal to the observed late-time expansion history. In our model, for dust ($w = 0$), the effective EOS parameter is
\begin{equation}
  w_{\rm eff}(H) = -\frac{6\sigma}{\kappa}H^2,
\end{equation}
so at the present epoch ($H=H_0$) one has
\begin{equation}\label{weff0_dust_relation}
    w_{\rm eff,0} = -\frac{6\sigma}{\kappa}H_0^2.
\end{equation}
On the other hand, the deceleration parameter in a flat FLRW universe~\cite{Baumann2022} filled with a single effective fluid is given by
\begin{equation}
  q = -\frac{1}{H^2}\frac{\ddot{a}}{a} = \frac{1}{2}\left(1+3w_{\mathrm{eff}}\right),
\end{equation}
so that we can immediately read off
\begin{equation}
  w_{\rm eff,0} = \frac{2q_0 - 1}{3}.
\end{equation}
Observationally, combined CMB+BAO+SNe analyses are consistent with an almost constant dark energy EOS parameter very close to a cosmological constant, \emph{e.g.} Planck 2018 finds $w_0 = -1.03\pm 0.03$ in a $w$CDM extension~\cite{Planck2018}. In a purely kinematical approach, parametrisations of the deceleration parameter constrained by SNe~Ia data show that the present universe is undergoing accelerated expansion, with a negative current deceleration parameter $q_0<0$~\cite{Cunha2008MNRAS}. Interpreting the late-time universe in our model as being described by a single effective dust-like fluid, we can equate the kinematical value of $w_{\rm eff,0}$ to the theoretical expression \eqref{weff0_dust_relation}, which gives
\begin{equation}\label{sigma_estimate}
  \sigma = -\frac{\kappa w_{\rm eff,0}}{6H_0^2}.
\end{equation}
Since $w_{\rm eff,0}<0$, this yields a positive coupling $\sigma>0$, as required in our accelerated branch. It is convenient to express this estimate in terms of the present critical density $\rho_{c,0} = 3H_0^2/\kappa$, for which \eqref{sigma_estimate} becomes
\begin{equation}
  \sigma\rho_{c,0} = -\frac{w_{\rm eff,0}}{2}.
\end{equation}
Thus, the dimensionless combination that controls the strength of the $RT$ coupling in the late universe is directly set by the observed effective equation-of-state parameter. For instance, taking a typical value $w_{\rm eff,0}\simeq -0.7$ one obtains $\sigma\rho_{c,0}\simeq 0.35$, \emph{i.e.}\ $\sigma$ is of order $\sigma\sim \mathcal{O}(1)\times\rho_{c,0}^{-1}$, consistent with the expectation that the non-minimal coupling becomes relevant only near the present critical density. In particular, the viability condition $1+\sigma T>0$ for dust with $T=-\rho$ requires $\sigma\rho<1$, which is comfortably satisfied at the present epoch since $\sigma\rho_{c,0}\simeq 0.35<1$. A more precise observational constraint on $\sigma$ would require a dedicated fit of the dimensionless combination $\sigma\rho_{c,0}$ to late-time cosmological data, such as $H(z)$, supernova and BAO measurements. However, the simple estimate \eqref{sigma_estimate} already shows that a positive $\sigma$ of order $\rho_{c0}^{-1}$ naturally reproduces the observed late-time effective EOS.

\subsection{Dust-dominated cosmology}

It is instructive to specialise the above results to the late-time matter era, modelled as dust ($w=0$), for which $T = -\rho$. The continuity equation \eqref{rho_continuity} then gives the usual scaling $\rho(a)=\rho_0(a_0/a)^3$. The algebraic relation \eqref{rho_of_H} between the physical density and the Hubble rate simplifies to
\begin{equation}\label{rho_of_H_dust}
  \rho(H) = \frac{3H^2}{\kappa-6\sigma H^2},
\end{equation}
valid as long as the denominator is positive. The Einstein-like Friedmann equations \eqref{Friedman} imply
\begin{equation}
  \rho_{\rm eff}^{(0)}(H) = \frac{3H^2}{\kappa},\qquad p_{\rm eff}^{(0)}(H) = -\frac{2\dot H+3H^2}{\kappa},
\end{equation}
and, using \eqref{peff2} with $w=0$, the effective pressure can be written purely as a function of $H$ as
\begin{equation}\label{peff_dust}
  p_{\rm eff}^{(0)}(H) = -\frac{18\sigma}{\kappa^2}H^4.
\end{equation}
The corresponding effective equation-of-state parameter, $w_{\rm eff}=p_{\rm eff}/\rho_{\rm eff}$, follows from \eqref{weff} and reads
\begin{equation}\label{weff_dust}
  w_{\rm eff}^{(0)}(H) = -\frac{6\sigma}{\kappa}H^2.
\end{equation}
Thus, for $\sigma > 0$, the effective dust fluid no longer behaves as pressureless matter. It develops a negative effective pressure. From \eqref{weff_dust} one sees that $w_{\rm eff}^{(0)}\simeq 0$ when $H^2$ is small, decreases to $-1/3$ at $H^2=\kappa/(18\sigma)$, and reaches $-1$ at $H^2=\kappa/(6\sigma)$ (see Table~\ref{tableEinsnone}). Hence the condition $w_{\rm eff}^{(0)}<-1/3$ marks the onset of accelerated expansion, $\ddot{a}>0$, while $\rho_{\rm eff}$ remains  positive. For dust, the energy conditions are most conveniently written as bounds on the dimensionless ratio $\sigma H^2/\kappa$, or equivalently on $H^2$ itself. For $\sigma>0$ one finds
\begin{equation}
  \text{NEC} = \text{WEC} = \text{DEC}: \qquad 0\leq H^2\leq \frac{\kappa}{6\sigma},
\end{equation}
which is equivalent to
\begin{equation}
  -1\leq w_{\rm eff}^{(0)}\leq 0.
\end{equation}
The SEC is more restrictive and requires
\begin{equation}
  0\leq H^2\leq \frac{\kappa}{18\sigma}, \qquad\Longleftrightarrow \qquad w_{\rm eff}^{(0)}\geq -\frac{1}{3}.
\end{equation}
Therefore the interval
\begin{equation}
  \frac{\kappa}{18\sigma}<H^2\leq \frac{\kappa}{6\sigma}
\end{equation}
is particularly important. Here, the NEC, WEC and DEC remain satisfied, whereas the SEC is violated, and the effective dust component drives an  accelerated expansion with
\begin{equation}
  -1\leq w_{\rm eff}^{(0)}< -\frac{1}{3}.
\end{equation}
There is, however, an additional viability requirement. For dust, $1+\sigma T>0$ is equivalent to $\rho<1/\sigma$, and using \eqref{rho_of_H_dust} this becomes
\begin{equation}
  H^2\leq H_\sigma^2= \frac{\kappa}{9\sigma}.
\end{equation}
This bound is stronger than the NEC/WEC/DEC limit, so the allowed range is reduced from
\begin{equation}
  0\leq H^2\leq \frac{\kappa}{6\sigma}
\end{equation}
to
\begin{equation}
  0\leq H^2\leq \frac{\kappa}{9\sigma}.
\end{equation}
The accelerating branch is therefore not removed, but narrowed to
\begin{equation}
  \frac{\kappa}{18\sigma} < H^2 \leq \frac{\kappa}{9\sigma},
\end{equation}
for which
\begin{equation}
  -\frac{2}{3}\leq w_{\rm eff}^{(0)}< -\frac{1}{3},
\end{equation}
and all energy conditions except the SEC are satisfied. To make the above discussion more concrete, it is useful to identify a few representative points in the viable accelerating dust branch. Using $w_{\mathrm{eff}}^{(0)}(H)=-6\sigma H^{2}/\kappa$ together with $q=(1+3w_{\mathrm{eff}})/2$, the onset of acceleration occurs at $H^{2}=\kappa/(18\sigma)$, where $w_{\mathrm{eff}}^{(0)}=-1/3$ and $q=0$. At the intermediate value $H^{2}=\kappa/(12\sigma)$, one obtains $w_{\mathrm{eff}}^{(0)}=-1/2$ and $q=-1/4$. Finally, the upper limit imposed by $1+\sigma T>0$, namely $H^{2}=\kappa/(9\sigma)$, corresponds to $w_{\mathrm{eff}}^{(0)}=-2/3$ and $q=-1/2$. Hence, the physically admissible $\sigma>0$ dust solutions describe a quantitatively quasi-de Sitter regime with moderate late-time acceleration, while exact de Sitter expansion is excluded by the viability bound. Throughout this interval, the NEC, WEC and DEC remain satisfied, whereas only the SEC is violated. For $\sigma<0$, the picture changes completely. Equation~\eqref{weff_dust} gives $w_{\rm eff}^{(0)}(H)\geq 0$, so the effective dust fluid becomes stiffer than ordinary dust and approaches a stiff fluid regime, $w_{\rm eff}^{(0)}\to 1$, as $H^2$ increases. The viability condition is now automatic. For dust one has $T = -\rho<0$, so $\sigma T > 0$ and hence $1+\sigma T>0$. In this branch, the NEC, WEC and SEC are always satisfied, while the DEC requires
\begin{equation}
  H^2\leq \frac{\kappa}{6|\sigma|},
\end{equation}
equivalently $w_{\rm eff}^{(0)}\leq 1$. Since $w_{\rm eff}^{(0)}\geq 0$, one also has $\rho_{\rm eff}+3p_{\rm eff}\geq 0$, and therefore
\begin{equation}
  \frac{\ddot a}{a}\leq 0.
\end{equation}
No accelerated expansion can occur in this branch. In summary, a positive coupling $\sigma>0$ turns dust into an effective fluid with negative pressure, allowing SEC violation and late-time acceleration over a finite interval of $H^2$ while preserving the NEC, WEC and DEC. By contrast, a negative coupling produces a stiffer but non-accelerating effective fluid.

\section{Conclusions}

In this work, we have provided a consistent cosmological analysis of the non–separable curvature–trace coupling $f(R, T) = R+\sigma RT$ in the framework of $f(R, T)$ gravity with a barotropic perfect fluid. Building on Brown's variational formulation of relativistic fluids and the corrected perfect-fluid equations, we have recast the modified field equations into an Einstein-like form with effective energy density, pressure, and an EOS parameter. This effective fluid description allows the NEC, WEC, SEC, and DEC to be written as simple algebraic inequalities involving the Hubble rate $H$, the equation-of-state parameter $w$, and the coupling constant $\sigma$. Specialising in a spatially flat FLRW universe, we have shown that radiation reproduces standard relativistic cosmology exactly since the trace $T$ vanishes, the $R\, T$ coupling becomes dynamically irrelevant, and all energy conditions are identically satisfied for any value of $\sigma$. The dust case displays a richer behaviour. 

For $\sigma > 0$, the effective dust acquires a negative pressure, with an effective EOS parameter that interpolates between standard dust at small $H^{2}$ and a cosmological constant-like behaviour as $H^{2}$ approaches its upper bound. We have identified a finite range of the Hubble parameter for which the NEC, WEC, and DEC remain satisfied while the SEC is violated. This is precisely the regime in which the effective fluid can drive accelerated expansion. For $\sigma < 0$, the effective dust becomes stiffer rather than softer, $w_{\rm eff}\geq 0$, and no accelerating phase is possible, although the DEC still imposes non-trivial bounds on the allowed Hubble range. An additional viability requirement in this model is the positivity of the effective gravitational coupling, $\kappa_{\rm eff}=\kappa/(1+\sigma T)$, which we enforce by demanding $1 + \sigma T > 0$. For barotropic fluids with $-1<w\leq 1/3$, this inequality constrains the physical density and, translated into the FLRW background via the Friedmann equations, yields an upper bound on $H^{2}$ that is, for $\sigma>0$, more restrictive than the bounds coming from the NEC, WEC and DEC. In the dust case, we have shown that the accelerating regime survives once this condition is imposed, but is confined to the window $\kappa/(18\sigma) < H^{2}\leq \kappa/(9\sigma)$, where $-2/3\leq w_{\rm eff} < -1/3$ and all conditions except the SEC are satisfied. Taken together, these results provide a transparent energy condition analysis of the non-trivial $R\,T$ coupling, in contrast to the large class of $f(R, T)$ models whose trace dependence can be absorbed into a redefinition of the fluid EOS and which therefore do not introduce new gravitational degrees of freedom for perfect fluids. We have also obtained an indicative estimate for the magnitude of the coupling $\sigma$ by matching the effective dust-like EOS parameter to kinematical determinations of the present deceleration parameter. At the background level this leads to a dimensionless combination $\sigma\rho_{c,0}$ of order unity, so that a coupling $\sigma\sim \mathcal{O}(1)\rho_{c,0}^{-1}$ can naturally reproduce the observed late time acceleration while satisfying the viability condition $1 + \sigma T > 0$ at the present epoch. A more refined constraint on $\sigma$ would require a full statistical analysis of cosmological data within the present model, including the redshift dependence of $H(z)$ and other background observables, as well as the growth of structure. Nonetheless, our simple estimate already shows that the coupling can act as an effective dark energy component at cosmological densities, without the need for exotic matter sources.

Several extensions of the present work suggest themselves. On the cosmological side, a natural extension is to move beyond the homogeneous background and study scalar, vector, and tensor perturbations to determine how the gravitational potentials evolve and how this affects observables such as the growth rate and the late-integrated Sachs–Wolfe effect~\cite{Sachs1967AJ}. This would allow one to confront the model directly with current and forthcoming large-scale structure and CMB data, and to examine possible degeneracies with phenomenological dark-energy parametrisations. It would also be of interest to relax the assumption of a single barotropic fluid and to consider mixtures of dust, radiation and a cosmological constant, as well as more general equations of state. A multi-fluid approach is of particular interest in this context, as one would choose only one fluid to be coupled non-minimally with the geometry, with the aim of this fluid component being a good dark matter or dark energy candidate. Finally, static, spherically symmetric solutions provide an independent setting for investigating curvature–trace couplings. It would be interesting to revisit the classical Buchdahl bound and to explore the maximum compactness of relativistic stars for the $R + \sigma RT$ model. Since the vacuum exterior remains Schwarzschild while the interior geometry is affected by the coupling, we expect the compactness limit to provide complementary, potentially more stringent constraints on $\sigma$. We leave this analysis for future work.

\appendix

\section{Constant-\texorpdfstring{$H$}{H} fixed points}
\label{app:constH}

We briefly discuss FLRW backgrounds with a positive constant Hubble parameter, $H(t)=\lambda > 0$, corresponding to de Sitter space or Minkowski space for $\lambda=0$, and clarify how such solutions fit into the effective fluid description used in the main text. As before, the field equations are
\begin{equation}
    \label{app:friedmann}
    3H^2=\kappa\rho_{\rm eff}\,,\qquad 
    2\dot H+3H^2=-\kappa p_{\rm eff}\,.
\end{equation}
Using $p_{\rm eff}=w_{\rm eff}\rho_{\rm eff}$ together with $\rho_{\rm eff}=3H^2/\kappa$ from the first equation, the second equation can be written as an autonomous evolution equation for $H$ which reads
\begin{equation}
    \label{app:Hdot}
    \dot H = -\frac{3}{2}\bigl[1+w_{\rm eff}(H)\bigr]H^2.
\end{equation}
Hence, the regular fixed points are
\begin{equation}
    \label{app:fixedpoints}
    H=0,
    \qquad\text{or}\qquad
    w_{\rm eff}(H)=-1\,.
\end{equation}
The case $H=0$ corresponds to a static Minkowski limit. Any solution with $\lambda\neq 0$ satisfies $w_{\rm eff}=-1$ at the effective level. Setting $H=\lambda\neq 0$ in~\eqref{app:friedmann} gives
\begin{equation}
    \label{app:dSweff}
    3\lambda^2=\kappa\rho_{\rm eff}\,,\qquad 3\lambda^2=-\kappa p_{\rm eff}
    \quad\Longrightarrow\quad
    p_{\rm eff}=-\rho_{\rm eff}\,,\qquad w_{\rm eff}=-1\,.
\end{equation}
Thus, independently of the microscopic matter content, a de Sitter background is possible if and only if the effective fluid behaves like a cosmological constant. For the barotropic Brown fluid $p=w\rho$ used in the main text, we recall our equation
\begin{equation}
    \label{app:weffH}
    w_{\rm eff}(H)=w+\frac{3\sigma}{\kappa}(3w+2)(3w-1)(1+w)H^2.
\end{equation}
Imposing $w_{\rm eff}(\lambda)=-1$ yields
\begin{equation}
    \label{app:weffminus1}
    w+\frac{3\sigma}{\kappa}(3w+2)(3w-1)(1+w)\lambda^2=-1
    \quad\Longleftrightarrow\quad
    (1+w)\left[1+\frac{3\sigma}{\kappa}(3w+2)(3w-1)\lambda^2\right]=0.
\end{equation}
There are two distinct cases to consider:
\begin{enumerate}
\item     
If $w=-1$, the matter density is constant, and a genuine de Sitter solution exists. Using the algebraic equation~\eqref{rho_of_H}$ for \rho(H)$, evaluated at $H=\lambda$, gives
\begin{equation}
    \label{app:rhodS}
    \rho_{\rm dS} = \frac{3\lambda^2}{\kappa+12\sigma\lambda^2},
\end{equation}
and the viability condition $1+\sigma T>0$ becomes
\begin{equation}
    \label{app:viabilitydS}
    1+\sigma T = 1-4\sigma\rho_{\rm dS}
    =\frac{\kappa}{\kappa+12\sigma\lambda^2}>0
    \quad\Longleftrightarrow\quad
    \kappa+12\sigma\lambda^2>0.
\end{equation}
In particular, for $\sigma>0$ this is automatically satisfied, and we have the general restriction $\sigma > - \kappa/(12\lambda^2)$.
\item 
If $w\neq -1$, then \eqref{app:weffminus1} requires
\begin{equation}\label{app:singularsurface}
    1+\frac{3\sigma}{\kappa}(3w+2)(3w-1)\lambda^2=0
    \quad\Longleftrightarrow\quad
    \kappa+3\sigma(3w+2)(3w-1)\lambda^2=0\,.
\end{equation}
However, this condition coincides with the degeneracy surface of the algebraic relation $\rho(H)$, see again~\eqref{rho_of_H}: the denominator in $\rho(H)$ vanishes, and the first Friedmann equation admits no solution with finite $\rho$ and $\lambda^2>0$. Equivalently, the apparent fixed point in~\eqref{app:fixedpoints} sits at a singular boundary of the reduced $(H,\rho)$ phase space. For completeness, solving~\eqref{app:weffminus1} for $w$ at fixed $(\sigma,\lambda)$ gives
\begin{equation}\label{app:wpm}
    w_\pm=\frac{1}{6}\left(-1\pm\sqrt{\,9-\frac{4\kappa}{3\sigma\lambda^2}\,}\right)
\end{equation}
(real for $\sigma<0$ or $\sigma\lambda^2\ge 4\kappa/27$), but these roots satisfy \eqref{app:singularsurface} and therefore do not correspond to regular cosmological solutions with finite matter density.
\end{enumerate}

\addcontentsline{toc}{section}{References}
\bibliographystyle{jhepmodstyle}
\bibliography{FRT}

\end{document}